\documentclass[aip,reprint]{revtex4-1}
\usepackage{graphicx}
\usepackage{dcolumn}
\usepackage{bm}
\usepackage{amsmath}
\usepackage{mathrsfs}
\usepackage{empheq}
\usepackage{url}
\usepackage{natbib}
\usepackage{aas_macro}
\usepackage{hyperref}
\usepackage[applemac]{inputenc}
\usepackage{color}

\begin{document}

\preprint{AIP/123-QED}

\title{Onset of fast ``ideal" tearing in thin current sheets: dependence on the equilibrium current profile}

\author{F. Pucci}
\email{fpucci@nifs.ac.jp, fpucci@princeton.edu}
\affiliation{National Institute for Fusion Science, National Institutes of Natural Sciences,
Toki 509-5292, Japan}
\affiliation{ Princeton Plasma Physics Laboratory, Princeton University, Princeton, New Jersey, USA}
\author{M. Velli}%
\email{mvelli@ucla.edu}
\affiliation{ University of California Los Angeles, USA
}%
\author{A. Tenerani}%
\email{annatenerani@epss.ucla.edu}
\affiliation{ University of California Los Angeles, USA
}%

\author{D. Del Sarto}%
\email{daniele.del-sarto@univ-lorraine.fr}
\affiliation{ Institut Jean Lamour UMR 7198 CNRS, Universit\'e de
Lorraine, Nancy, France
}%

\date{\today}

\begin{abstract}
In this paper we study the scaling relations for the triggering of the fast, or ``ideal", tearing instability starting from equilibrium configurations relevant to astrophysical as well as laboratory plasmas that differ from the simple Harris current sheet configuration. We present the linear tearing instability analysis for equilibrium magnetic fields which a) go to zero at the boundary of the domain and b) contain a double current sheet system (the latter previously studied as a cartesian proxy for the m=1 kink mode in cylindrical plasmas). More generally, we discuss the critical aspect ratio scalings at which the growth rates become independent of the Lundquist number $S$, in terms of the dependence of the $\Delta'$ parameter on the wavenumber $k$ of unstable modes. The scaling $\Delta'(k)$ with $k$ at small $k$ is found to categorize different equilibria broadly: the critical aspect ratios may be even smaller than $L/a \sim S^{\alpha}$ with $\alpha=1/3$ originally found for the Harris current sheet, but there exists a general lower bound $\alpha\ge1/4$.
\end{abstract}

\pacs{Valid PACS appear here}
\maketitle

\section{Introduction}
It is widely accepted that the mechanism responsible for explosive events, both in space and laboratory plasmas, is magnetic reconnection, a complex phenomenon that involves several scales and physical effects in which the dynamical evolution is affected by the initial configuration. One way in which reconnection may be triggered is via the tearing instability of thin current sheets. The latter is characterized by a fundamental parameter $\Delta'$, which depends uniquely on the equilibrium magnetic field through the current density and its gradient, as well as the specific boundary conditions to which the plasma is subject. The $\Delta'$ parameter is also important for the subsequent non-linear evolution of magnetic islands in different geometrical configurations \citet{Furthetal:1973, Connoretal:1988, 2004PPCFOtt}, though $\Delta'$ on its own does not suffice to fully determine the dynamics of the system \citet{Militelloetal:2014}.
Studying the instability of a Harris current sheet, \citet{PucciVelli2014} (heareafter PV14) showed that for thin current sheets of thickness $2a$ and length $2L$ (assumed to be of the same order as the macroscopic length scale of the system), the specific dependence of the inverse aspect ratio $a/L$ on the Lundquist number $S$  determines when the instability becomes fast or ``ideal", where by the latter term it is implied that the growth rate no longer depends explicitly on the Lundquist number. In other words, PV14 showed that the $S\rightarrow \infty$ limit of resistive MHD is not ideal MHD, but rather a magnetohydrodynamics where the frozen in condition breaks down in thin current sheets on ideal timescales $\tau \sim L/v_A$, with $v_A$ the characteristic Alfv\'en speed. This suggests that reconnection survives in the limit $S\rightarrow \infty$ but that no divergence of growth rates occurs, the latter result being implicated by the studies of the plasmoid instability (\citet{loureiro07, HuangBattacharjee2016}), 
based on the Sweet-Parker current sheet and corresponding island number scaling laws. More specifically, by assuming the form $a/L\sim S^{-\alpha}$, PV14 showed that the critical inverse aspect ratio for current sheets was reached at $\alpha=1/3$, at which point the fastest growing tearing mode has a time-scale comparable to the Alfv\'en crossing time and, more fundamentally, is independent of $S$.  Therefore, the value $1/3$ separates slowly unstable current sheets (i.e. with growth rate scaling as a negative, fractional exponent of the Lundquist number) from those so violently unstable (with a growth rate scaling as a positive exponent of the Lundquist number, including the famed Sweet-Parker configuration) that they will never form in the first place. Simulations of a collapsing current sheet by \citet{Teneranietal2015b} demonstrated the fundamental correctness of the above scenario for a configuration comprised of an isolated current sheet of the Harris type. 

PV14 also discussed a number of other consequences of  ``ideal" tearing: namely that the singular layer of the tearing mode has a thickness $2\delta$ which scales as $\delta/L \sim S^{-1/2}$ and a number of islands born out of the linear instability that scales as $kL \sim S^{1/6}$. The corollary of the result that at $a/L \sim S^{-1/3}$ the fastest growing mode no longer scales with $S$ is that such modes, close to the maximum growth rates, remain the only growing modes  at asymptotically large $S$ , as modes at any other values of the wave number end up decaying in time. Such result is now corroborated by other simulations (Huang et al., 2017), which include the effects of flows and the dynamics of current sheet thinning, even though flows are known to play a stabilizing effect at low Lundquist numbers (Chen et al. 2018, Bulanov et al. 1978). 

The scaling in analysis in PV14 was fundamentally dependent on the details of the chosen equilibrium profile, the Harris current sheet with characteristic gradient scale $a$. On the other hand isolated Harris-type current sheet are probably a rare occurrence in natural plasmas, where magnetic configurations more complex than the planar sheets are continuously perturbed by background turbulence and/or by macroscopic forcings, which may drive the field to ideal instability before reconnection takes place. Multiple current sheet configurations commonly occur in astrophysical plasmas: rapid rotators with an inclined magnetic axis, such as the case of pulsar magnetospheres \citet{Contopoulos2007}, lead to a waving current sheet which piles up into a series of alternating current layers almost orthogonal to the rotation axis. In the solar corona, the helmet streamers where the closed field is pinched off into the solar wind, are not always completely dipolar, but may contain multiple polarities, leading to the formation of multiple current sheets as the streamer is stretched into the wind \citet{KleinBurlaga1980,Woo1994,Crookeretal1993,DalhburgKarpen1995}.
Multiple current sheet configurations have been observed in laboratory plasmas as well: as pointed out e.g. by \citet{Onoetal1997}, both for Reversed Field Pinches and Tokamaks, the periodic and/or continuous self-organization of their magnetic configurations produces multiple reconnection points \citet{Longecopeetal2002,Bierwageetal2005}. 

In addition, recent attempts incorporate the ``ideal" tearing concept in models of fully developed MHD turbulence \citet{2017PhRvL.118x5101L, 2017MNRAS.468.4862M}: in these calculations the scaling relations for current sheet stability modify the energy spectral index of the cascade. But the current sheets arising in a turbulent cascade most likely have a whole spectrum of current profiles crossing them, so that modifications of the critical aspect sheet ratio dependencies arising from the detailed current profiles will also end up affecting the energy cascade. We will further comment on this latter aspect in the conclusion, but it is clearly of great interest to extend the stability study to more general configurations with differing magnetic field profiles. 

Understanding the dependence of the classic tearing mode instability on the structure of any given equilibrium depends (at least assuming free or distant boundaries) on one fundamental parameter, the so called \textit{delta prime} or  $\Delta'$. The latter is determined by resolving the first order perturbed momentum equation across the current sheet in the ideal limit.  The perturbed, reconnecting magnetic field component (which will be called $b$) is found to be continuous across the sheet, but its first derivative retains a finite jump, (that will be ultimately resolved by including resistive diffusion in the calculation):
\begin{equation}
\Delta' = a\,  {\rm lim}_{\epsilon \rightarrow 0 } {b(+\epsilon) - b(-\epsilon) \over b(0)};
\end{equation}
here $b$ depends only on the coordinate across the sheet and $a$ represents the typical equilibrium scale-length (or current sheet half-thickness). 

The parameter $\Delta'$ depends on the normalized wave-number $ka$ (along the sheet) of the reconnecting velocity and magnetic field perturbations. In this paper we investigate how, in the limit $ka <<1$, different dependencies $\Delta'(k)$ determine the corresponding ``ideal tearing" scaling thresholds, i.e. the aspect ratio scaling index $\alpha$ in $a/L \sim S^{-\alpha}$, at which the maximum growth rate no longer depends on $S$. 
This affects not only the critical aspect ratio, but also the number of islands that emerge from the linear phase once the critical aspect ratio is reached and the tearing instability becomes fast. 

We will present detailed solutions for two cases: a configuration where the magnetic field far from the sheet vanishes and a double current sheet configuration. The plan of the paper is the following. In the next section we recall some terminology and properties related with the concept of the "ideal" tearing (IT hereafter) instability and we describe it in terms of the parameters that define the equilibrium configurations. In section three we discuss the scaling relations for an equilibrium magnetic field which goes to zero at the boundary of the domain, recovering the trigger condition for the "ideal" tearing mode to be developed. In section four we will discuss the transition to different regimes of a double current sheet configuration. Finally in the last section we discuss the results and the more general relevance to nonlinear plasma dynamics.  

\section{Instability regimes in slab tearing reconnection.}\label{sec:II} 
Consider a general slab magnetic field configuration described as:
\begin{equation} 
\label{eq:equilibriumMAGFILD}
{\bf B}=B_{0} \,\mathrm{F}(y/a)\,{\bf e}_{x} + B_{0z}(y/a)\, {\bf e}_{z},
\end{equation}
where $F(y/a)$ is a function of the coordinate $y$, $a$ defining the scale of the gradient of magnetic field and $B_{0z}$ is a field which may depend on $y/a$ to guarantee equilibirum.  We study the linearized resistive MHD equations, where the frozen in condition for magnetic field lines is locally violated due to the effect of a finite magnetic diffusivity $\eta$; we neglect kinetic scale effects in Ohm's law (see \citet{DelSartoetal2016,PVT17} for the extension of the ``ideal" tearing scenario including such terms in Ohm's law). Quantities are assumed to be uniform in the direction perpendicular to the plane of the equilibrium magnetic field, i.e. $\partial /\partial z=0$.
The equations are non-dimensionalized by normalizing the magnetic field and its perturbation in terms of the mean field $B_0$, the wavenumber $k$ along $x$ with $a$ and we introduce the non-dimensional displacement $\xi= iv_y/(\gamma a)$ and $b=b_y/B_0$. The growth rate is normalized to the macroscopic Alfvén time as previously defined, $\tau_A = L/v_A$ where the Alfvén speed $v_A = B_0/\sqrt{4\pi\rho}$ ($\rho$ is the plasma density); the Lundquist number is defined as $S=L v_A/ \eta$. We therefore define a new independent variable $\bar{y} = y/a = (y/L)\,S^{\alpha}$, and denote derivatives with respect to $\bar{y}$ with $'$.  We will consider an equilibrium current sheet aspect ratio scaling as $a/L \sim S^{-\alpha}$. The tearing mode equations become:
\\[0.08ex]
\begin{subequations}
\footnotesize
\label{tearingvelli3}
\begin{empheq}[left=\empheqlbrace]{align}
\label{tearingvelli4}
&{\tilde{\gamma}}^{2}\,S^{-2\alpha}( {\xi}''-{k^{2}} {\xi})=
-{k}\left[ {F}(\bar{y})\left( {b}''-{k}^{2}{b}\right)-{F}''(\bar{y}){b}\, \right] \\[2ex]
 \label{tearingvelli5}
&{b}={k}{F}(\bar{y}){\xi}+\,{1\over{\tilde{\gamma}} }S^{(2\alpha-1)}\left({b}''-{k}^{2}{b}\right).
\end{empheq}
\end{subequations}
\\[0.2ex]
where $\tilde{\gamma}=\gamma \tau_A$ is the non dimensional growth rate of the instability. To recover the classic tearing results, in which $a=L$, $\alpha=0$ while the Lundquist number and the Alfvén speed must be replaced with $\bar{S}=a\, v_A/\eta$ and $\bar{\tau}_A=a/v_A$ respectively.
The asymptotic, large Lundquist number analysis of the solutions to the above equations, subject to the boundary conditions that the velocity and magnetic field perturbations vanish far from the sheet, show that the solution consists of two regions: a boundary layer of thickness $2\delta$ around the center of the current sheet (which we will center around $y=0$), and an outer regions where diffusion may be neglected. Asymptotic matching then requires the jump in derivative of $b$ as calculated from the outer solution only, the $\Delta'$ parameter defined above, to equal the jump in derivative of $b$ calculated moving outwards from the inner solution.

The continuous spectrum of tearing unstable modes is then described in the two limits of large $\Delta'$ and small $\Delta'$ regimes (also known as the constant-$\psi$ ordering), in terms of whether $\Delta' \delta/a \gg 1$ or $\Delta' \delta/a \ll 1$. The maximum growth rate can be found by matching the two regimes, i.e. where $\Delta' \delta/a\simeq 1$, \citet{Bhattacharjeeetal2009, lou_2013, DelSartoetal2016, Teneranietal2016}.  We recall the results (where we have defined $\delta \equiv \delta/a$)
\small
\begin{equation}
\label{eq:LD}
\gamma_{_{LD}}\bar{\tau}_A\simeq k^{\frac{2}{3}}\bar{S}^{-\frac{1}{3}}\qquad \delta_{_{LD}}\sim (\bar{S}k)^{-\frac{1}{3}}
\end{equation}
\begin{equation}\label{eq:SD}
 \gamma_{_{SD}}\bar{\tau}_A\simeq A^{\frac{4}{5}}k^{\frac{2}{5}}(\Delta')^{\frac{4}{5}}\bar{S}^{-\frac{3}{5}}\qquad \delta_{_{SD}}\sim 
(\bar{S}k)^{-\frac{2}{5}}(\Delta')^{\frac{1}{5}}
\end{equation} 
\normalsize
\\
where $A\equiv \Gamma(1/4)/(2\pi\Gamma(3/4))$ is a coefficient given in terms of the values of Euler's $\Gamma$-function.
If we introduce a dependence of $\Delta'$ on the wavenumber $k$ at small $k$ as a power law, and assume this to hold also at the
maximum growth wave-number $k_{_M}$,
\begin{equation}
\label{eq:Delta_k_M}
\Delta'(k_{_M})\sim C k_{_M}^{-p},
\end{equation} 
where $C$ and $p$ are specified by the magnetic field equilibrium profile,
we can introduce this into the small $\Delta'$ dispersion relation regime, \citet{DelSartoetal2016}. Assuming that the two regimes are correct at maximum growth,
using Eq.(\ref{eq:Delta_k_M}-\ref{eq:SD}) and  $\gamma_{_{SD}}\simeq\gamma_{_{LD}}$  we find
\begin{equation}
\begin{array}{r@{}l}
\label{eq:Delta_k_M2}
&k_{_M}=C^{\frac{3}{1+3p}}A^{\frac{3}{1+3p}}\bar{S}^{-\frac{1}{1+3p}}\qquad \\[2ex]
&\gamma_{_M} \bar{\tau}_A \simeq C^{\frac{2}{1+3p}}A^{\frac{2}{1+3p}}\bar{S}^{-\frac{1+p}{1+3p}} 
\end{array}
\end{equation}
\normalsize
and for the singular resistive layer $\delta(k_{_{M}})\sim \bar{S}^{-\frac{p}{1+3p}}$.  For the classical Harris-pinch equilibrium, plotted in Fig.\ref{Fig:different_eq}, the estimation $p\simeq 1$ obtained in the instability $k$ range of the corresponding $\Delta'(k)$ reveals to be quite reasonable, providing the known results  $k_{_M}\sim S^{-{1}/{4}}$ and $\tilde{\gamma}_{_M}\sim S^{-{1}/{2}}$ \citet{FKR}, verified numerically in \citet{PucciVelli2014}. As discussed in \citet{LandidelZAetal2015}, even in the presence of the general equilibrium in Eq.\ref{eq:equilibriumMAGFILD} it is easy to show that the linear analysis of the instability is unchanged with respect to \citet{PucciVelli2014}.

The relationships in Eq.\eqref{eq:Delta_k_M2} can be generalized to a current sheet characterized by a thickness $a$ and a length $L$ and in particular, since $S= v_A L/\eta$ and $\tau_A= L/v_A$, we have $k_{_M}L/a\sim S^{-{1}/{4}}(a/L)^{-5/4}$ and $\gamma_{_M}\tau_A\sim S^{-{1}/{2}}(a/L)^{-3/2}$ \citet{PucciVelli2014} \citet{Teneranietal2016}. While the IT condition $\gamma \tau_A \sim 1$ separates slowly unstable current sheets from \textit{supertearing} instabilities \citet{loureiro07, HuangBattacharjee2016}, the condition to have \textit{plasmoids}, i.e. more than one island along the current sheet length associated to the fastest mode, since $k_{_M}L/a=n\,\pi$, is having an integer $n>1$. This means $S^{-{1}/{4}}(a/L)^{-5/4} >  \pi$, i.e. for an inverse aspect ratio $a/L \sim S^{-\alpha}$
\begin{eqnarray}
\alpha = \dfrac{1}{5} \left( 4 \dfrac{\mathrm{log}_{10} \, \pi}{\mathrm{log}_{10}  \,  S} +1\right).
\label{eq:condition_plasmoid}
\end{eqnarray}
i.e. for instance if $S=10^{12}$, $\alpha \ge 0.233$.


\section{Boundary Magnetic Field}\label{sec:III}
\begin{figure}
\includegraphics[width=80mm]{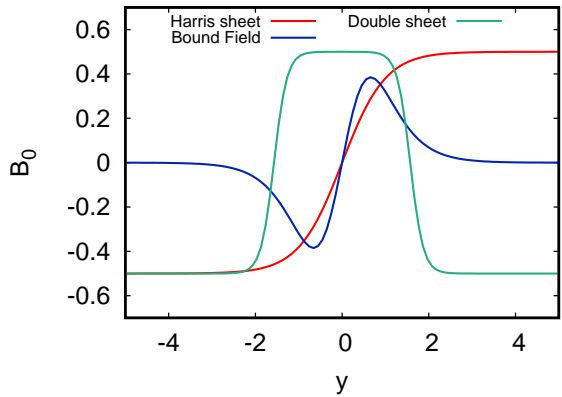}%
\caption{Typical profiles of the different magnetic field equilibria considered here: Harris equilibrium (red), magnetic field which goes to zero at the boundary of the domain (blue) and a double current sheet configuration (green).}
\label{Fig:different_eq}
\end{figure}

\begin{figure}
\centering
\includegraphics[width=83mm]{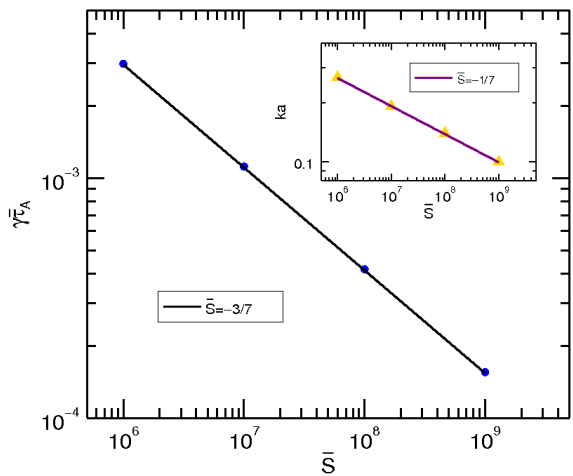}%
\caption{Scaling of the maximum growth rate as a function of the Lundquist number and wave vector (in the inset) corresponding to the maximum growth rate for the classic tearing mode, i.e. inverse aspect ratio $a/L=1$, for an equilibrium profile of the form Eq.\eqref{eq:equilibrio2}.}
\label{scalings_eq2}
\end{figure}

\begin{figure}
\includegraphics[width=83mm]{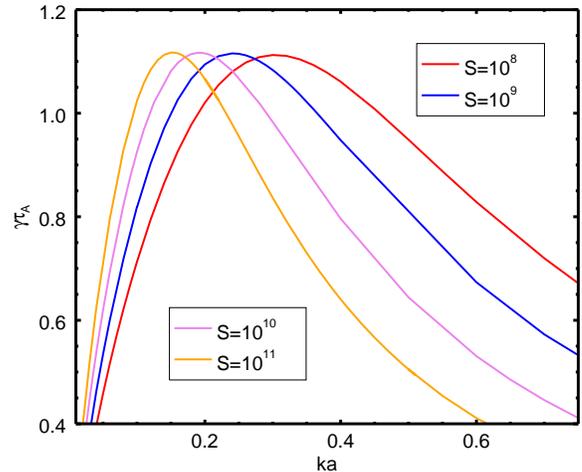}
\caption{\textit{Growth rate normalized to the Alfvén time ($\tau_A=L v_A/\eta$) as a function of the wave vector for a current sheet with an inverse aspect ratio $a/L \sim S^{-3/10}$, for an equilibrium profile of the form Eq.\eqref{eq:equilibrio2}. Notice the maximum growth rate does not depend on the Lundquist number and is of the order of an Alfvén time.}}
\label{eq2_03}
\end{figure}

Consider an equilibrium magnetic field of the form 
\begin{equation} 
\label{eq:equilibrio2}
{\bf B}=B_{0} \dfrac{\mathrm{tanh}(y/a)}{\mathrm{cosh}^2(y/a)}\,{\bf e}_{x},
\end{equation}
showed in Fig.\ref{Fig:different_eq}, where the equilibrium magnetic field is localized in the $y$ direction. 
For this equilibrium magnetic field profile, the $\Delta'$ parameter is 
\begin{eqnarray}
\Delta'=2\left[ \dfrac{6\kappa^2-9}{\kappa(\kappa^2-4)}-\kappa \right]
\label{eq:deltaprime_eq2}
\end{eqnarray}
where $\kappa=k^2+4$ and the instability regime occurs for $k<\sqrt{5}$, \citet{Porcellietal:2002}. For $k\ll1$ Eq.\eqref{eq:deltaprime_eq2} can be express in the form Eq.\eqref{eq:Delta_k_M}, for $p=2$. From Eq.\eqref{eq:Delta_k_M2}, for the maximum growth rate we find
\begin{eqnarray}
\label{eq:Delta_k_M3}
&& k_{_M}=C^{\frac{3}{7}}A^{\frac{3}{7}}\bar{S}^{-\frac{1}{7}}\qquad \\[2ex]
&& \gamma_{_M}\simeq C^{\frac{2}{7}}A^{\frac{2}{7}}\bar{S}^{-\frac{3}{7}} 
 \end{eqnarray} 
which is verified numerically as shown in Fig.\ref{scalings_eq2}.
Applying the renormalization criterion, as described in Sect.\ref{sec:II}, neglecting the dependence on the coefficients $A$ and $C$, one gets
\begin{equation}
\gamma \tau_A \approx S^{-3/7}  \left( \dfrac{a}{L}\right)^{-10/7},
\label{eq:fastest_doubleIT}
\end{equation}
and assuming a scaling for the inverse aspect ratio $\dfrac{a}{L} \sim S^{-\alpha}$, the condition 
\begin{equation}
\gamma \tau_A \approx 1 \Rightarrow \alpha=3/10.
\label{eq:alpha_index_double}
\end{equation}
As shown in Fig.\ref{eq2_03} for this value of $\alpha$ for $S\rightarrow \infty$ the growth rate reaches an asymptotic value of $\gamma \bar{\tau}_A \sim 1.1$, independent of the resistivity. For arbitrary $p$, Eq. \eqref{eq:alpha_index_double} generalizes to 
\begin{equation}
\alpha=\dfrac{1+p}{2(1+2p)},
\label{eq:alpha_index_generic}
\end{equation}
so $0.25 \le \alpha \le 0.5$. 
Notice that $a/L \sim S^{-3/10} > S^{-1/3}$, i.e. the critical inverse aspect ratio is larger in this case with respect to an initial Harris type current sheet, so still much thicker than the Sweet Parker current sheet configuration. 

\section{\label{sec:IV} The double tearing mode.}
\begin{figure}
\centering
\includegraphics[width=83mm]{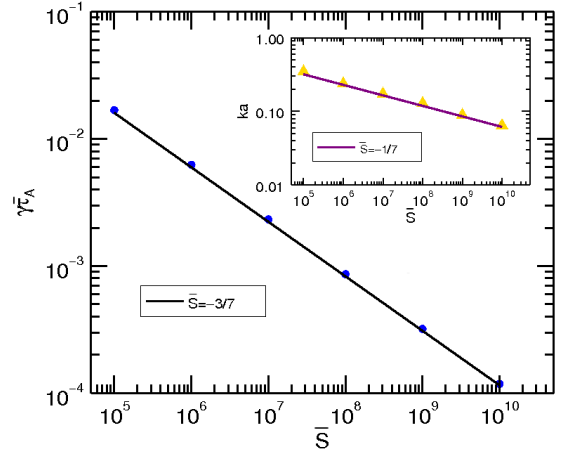}
\caption{Scaling of the maximum growth rate as a function of the Lundquist number and wave vector corresponding to the maximum growth rate for the configuration described by Eq.\eqref{Eq:double_equilibrium}.}
\label{Fig:double_maximum_classic}
\end{figure}

\begin{figure}
\centering
\includegraphics[width=83mm]{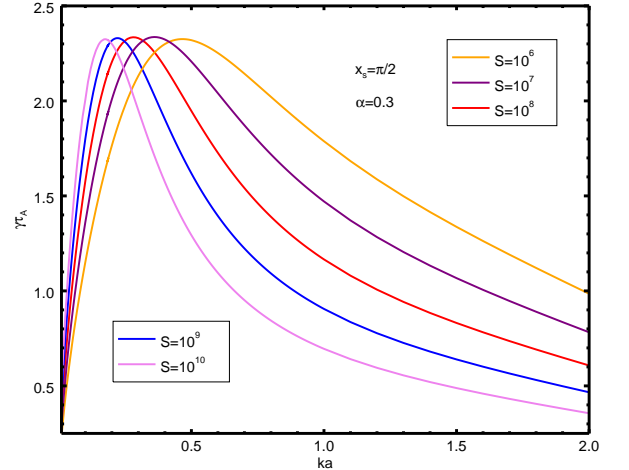}
\caption{Growth rate for the configuration described by Eq.\eqref{Eq:double_equilibrium}, as a function of the wave vector for $\alpha=3/10$. Different colors code different values of the Lundquist number.}
\label{Fig:ideal_1}
\end{figure}

Consider now a double current sheet configuration, described by
\begin{equation}
{\bf B}= \dfrac{B_0}{2} \Bigl[{\rm tanh}\left({y-x_s\over a}\right) -{\rm tanh}\left({y-3x_s\over a}\right)-1 \Bigr] {\bf e} _x, 
\label{Eq:double_equilibrium}
\end{equation}
where $x_s$ is the semi-distance between the two rational surfaces and $a$ the thickness of the current sheet, see Fig.\ref{Fig:different_eq}. The mutual interaction between the rational surfaces may lead to a coupled tearing mode so that, depending on the distance which separates the two current sheets, the dispersion relation presents two asymptotic regions \citet{Pritchett1980}. For the classic double tearing mode instability the latter paper shows that, setting $\bar{x}_s=x_s/a$, when
\begin{equation}
(k/\bar{S})^{1/3} \ll k \bar{x}_s\ll (k^2/\bar{S})^{1/9}, 
\label{Eq:interval_validity}
\end{equation}
the growth rate scales as 
\begin{equation}
\gamma \bar{\tau}_A \approx (k^2 B'^2_0/\bar{S})^{1/3}.
\label{eq:fastest_double}
\end{equation}
The lower limit in Eq.\eqref{Eq:interval_validity} guarantees that $\delta \ll \bar{x}_s$, i.e. the inner resistive layer is much smaller than the distance between the two rational surfaces, which allows the matching between the inner resistive region and the outer ideal region. The upper limit is provided by $|\lambda_h| \ll 1$ where $|\lambda_h| := |\gamma_A \bar{\tau}_A (\bar{S}/ (\hat{k} B'_0)^2)^{1/3}|$, where $\gamma_A \bar{\tau}_A$ is the double kink mode growth rate, described by Eq.(17) of \citet{Pritchett1980}. The latter depends on the equilibrium configuration, in particular for Eq.\eqref{Eq:double_equilibrium}:
\begin{eqnarray}
&&\gamma_A \bar{\tau}_A = -\left(\dfrac{\pi k^3}{B'_0(\bar{x}_s)}\right) \int^{\bar{x}_s}_0 \ \mathrm{tanh}^2\left(\dfrac{x'-x_s}{a}\right) dx' \nonumber \\
&&=\pi k^3  \bar{x}_s^3,
\label{Eq:double_kink_mode_growth}
\end{eqnarray}
i.e. $\gamma_A \bar{\tau}_A\sim k^3 \bar{x}_s^3$, so that $|\lambda_h| \sim 1$ provides the right inequality in Eq.\eqref{Eq:interval_validity}.
In the opposite case, for $|\lambda_h| \gg 1$
\begin{equation}
(k^2/\bar{S})^{1/9} \ll k \bar{x}_s \ll 1, 
\end{equation}
the growth rate scales as
\begin{equation}
\gamma \bar{\tau}_A  \approx \left(\dfrac{8\,\Gamma(5/4)}{\gamma_A \bar{\tau}_A\Gamma(-1/4)}\right)^{4/5}(k^2 B'^2_0/\bar{S})^{3/5},
\label{eq:small_deltaprime}
\end{equation}
where, $\Gamma$ is the Euler gamma function. The upper limit is necessary to provide the solution in the ideal region for the asymptotic matching. The transition between the two regimes, for $B'_0=1$, occurs for $ k \sim \bar{x}_s^{-9/7} \bar{S}^{-1/7}$. The regime identified by Eq.\eqref{eq:fastest_double} appears to be the fastest one, in particular for a fixed distance between the two resonant surfaces, we have the maximum growth rate at the transition between the two regimes, i.e. for 
\small
\begin{equation}
\gamma \bar{\tau}_A \approx (\bar{x}_s^{-9/7} \bar{S}^{-1/7})^{2/3} \bar{S}^{-1/3} \Rightarrow \gamma \bar{\tau}_A \approx \bar{x}_s^{-6/7} \bar{S}^{-3/7} .
\label{eq:fastest_double2}
\end{equation}
\normalsize
We verified numerically the scaling for the maximum growth rate as a function of the Lundquist number, as well as the wavevector corresponding to the maximum growth rate, as shown in Fig.\ref{Fig:double_maximum_classic}\\
If we want to generalize the IT condition for the double current sheet configuration we must take into account the condition $k\, \bar{x}_s \ll1$, so 
\begin{equation}
\gamma \tau_A \approx \bar{x}_s^{-6/7} S^{-3/7}  \left( \dfrac{a}{L}\right)^{-10/7},
\label{eq:fastest_doubleIT}
\end{equation}
i.e. the distance between the two resonant surfaces is fixed and of the order of $a$. Assuming a scaling for the inverse aspect ratio $\dfrac{a}{L} \sim S^{-\alpha}$, the condition 
\begin{equation}
\gamma \tau_A \approx 1 \Rightarrow \alpha=3/10,
\label{eq:alpha_index_double2}
\end{equation}
verified numerically, as shown in Fig.\ref{Fig:ideal_1}.
In terms of wavevector the coupled tearing mode regime occurs for
\begin{equation}
kL \left( \dfrac{a}{L}\right) \ll \bar{x}_s^{-9/7} S^{-1/7} \left( \dfrac{a}{L}\right)^{-1/7},
\end{equation}
Using \eqref{eq:alpha_index_double2}:
\begin{equation}
kL \ll \bar{x}_s^{-9/7} S^{12/35},
\label{eq:condition_on_K}
\end{equation}
where $\bar{x}_s=x_s/a=\pi/2$ in our case. Since the minimum available wavevector is $k_{min}L =  \pi$, for $ S \rightarrow \infty$ the coupled tearing regime is realized for a large number of modes (for instance $\bar{S}=10^7$ the condition holds $\pi \ge kL \ll 140$). The Eq.\eqref{eq:condition_on_K} can be interpreted as a condition on the distance between the two rational surfaces in order to have a coupled tearing mode regime, i.e $\bar{x}_s \ll (1/2\pi)^{7/9} S^{4/15}$. For $x_s/a \gg 1$, for instance in the case of $x_s/L \sim 1$, the analysis in \citet{Pritchett1980} is invalid yet a coupled configuration should still exist. Indeed we estimated numerically the value of $\alpha$ in the case $x_s/a \gg 1$ finding that $3/10 \le \alpha \le1/3$.

\section{Conclusions}
In this paper we have discussed the critical aspect ratios required for the fastest tearing mode growth times to reach values comparable to ideal time-scales for current sheet equilibria differing from the Harris current sheet. Under fairly general conditions, it was shown that the crucial parameter affecting the
scaling of the critical aspect ratio with the Lundquist number was the power-law scaling of the $\Delta'$ parameter in the large $\Delta'$ regime, $\Delta' \sim k ^{-p}$. For the case of a single current sheet profile, such as the one presented in Sect.\ref{sec:III}, deviations from the $p=1$ limit of Eq.(\ref{eq:Delta_k_M}), modify the scaling relations of the classic tearing mode instability resulting ins modification of the trigger coefficient $\alpha$, constraining it to be in the range $0.25 \le \alpha \le 0.5$, i.e. with a thickness of the limiting current sheet greater than the Sweet Parker sheet type. 

In the case of a double layer configuration, the distance between the two resonant surfaces, determines wether the tearing mode instability develops in a coupled or decoupled regime, while the maximum growth rate occurs at the transition between the two. The rescaling argument provides a value for the trigger aspect ratio $a/L \sim S^{-3/10}$ very similar to the value $9/29$ estimated in the linear numerical solutions presented in \citet{2017ApJ...837...74B}. 

As discussed in \citet{Teneranietal2015b,Teneranietal2016} the critical aspect ratio criterion for fast reconnection is also important because it affects the subsequent nonlinear evolution of a current sheet or current sheet system in the nonlinear phase. More specifically, the criterion can be applied to understand the trigger of nonlinear secondary reconnection and the subsequent self-similar collapse, at least in 2D, giving insight into turbulence initiated by tearing in an initially laminar current sheet. With application to  
the secondary reconnecting instabilities in the nonlinear regime of the m=1 resistive kink mode in tokamaks, \citet{DDSOTTAV} have discussed the nonlinear transition to fast reconnection stressing the re-scaling (or embedding) effects due to the decreased magnetic field on the reconnecting current sheet compared to the overall global field intensity. They found subsequent non-linear secondary instability to evolve in this case on a very weakly $S$ dependent time-scale.

Recently, attempts have also been made to incorporate the effects of the fast tearing of thin current sheets into models of fully developed MHD turbulence (\citet{2017PhRvL.118x5101L}, \citet{2017MNRAS.468.4862M}). In these models, the authors assume - as has been often conjectured and been previously shown in many numerical simulations (see, e.g.,  the simulations of \citet{Rappazzo:2008} and \citet{2017PhFl...29c5105Y}) - that the fundamental, dissipative coherent structure evolving out of the anisotropic nonlinear cascade is a current sheet (with aspect ratios depending on the scale and precise nature of the anisotropy). The idea is that the subsequent destabilization of the current sheet, once its aspect ratio has become too large, affects the subsequent nonlinear cascade and therefore also the inertial range spectrum. In this case, the trigger condition is given by the fastest tearing mode growth time becoming comparable to the nonlinear transfer time at the given scale (both time-scales  independent of the Lundquist number). This gives a condition on the scale at which the inertial range spectrum is modified:  the actual aspect ratio at which the tearing mode loses its scaling with the Lundquist number therefore also affects the subsequent inertial range spectrum. The results of the present paper are relevant in that they bracket the ranges of critical aspect ratios for a specific set of current profiles.  It is important to note that in the turbulent cascade additional details will almost certainly become relevant, from the combined flow-current profile of the individual current sheet (coherent structure), to the precise topological nature of the structure itself (neutral vs non-neutral current sheets, for example). Finally, though the trigger aspect-ratios are fundamental in the 2D nonlinear evolution, a 3D system may evolve directly through ideal secondary instabilities permitted by the initial reconnection-determined topology change: for example, kink-type instabilities may take over. This would modify the nonlinear recursive resistive evolution discussed by \citet{Teneranietal2015b}. We plan to address some of these issues in upcoming papers.

\section{Acknowledgements}
This work was supported by the NASA Parker Solar Probe Observatory Scientist grant  NNX15AF34G and the NSF-DOE Partnership in Basic Plasma Science and Engineering Award 1619611. The research presented also benefited greatly from discussions at the ISSI team meetings on “Explosive Processes in the Magnetotail: Reconnection Onset and Associated Plasma Instabilities” held in Bern, Switzerland from 
17–21 October 2016 and 23-27 October 2017.

\bibliography{Bib}

\end{document}